# Quantum-classical neurons upgraded with optical spin qubits for advanced computing and networking architectures

**Authors:** Osama M. Nayfeh[1†]

**Affiliations:** [1]Basic and Applied Research Division, Code 71780, Cyber/Science and Technology Department, Naval Information Warfare Center (NIWC) Pacific, Naval Base Point Loma, San Diego, CA USA

[†]**Corresponding author:** osama.m.nayfeh.civ@us.navy.mil; (619) 553-2770

## Abstract

Hardware neurons incorporating built-in memory components are critical technologies for implementing large scale neural networks that exhibit adaptive-itinerant behavior and produce biological-inspired spiking patterns that match neuroscience known operations *(1)*. Moreover, neurons capable of full quantum information processing through their qubit states and quantum trajectory provides capability for hybrid quantum-classical operations through the level of coherence/entanglement and non-Markovianity *(2,3)*. These attributes further contribute to solving complex quantum mechanical problems by providing a rich and diverse group for expressing quantum neural state*s*. Upgrading these hardware neurons now with optical spin qubits excitable with pulsed laser excitation in the near infrared (NIR) and designed from negatively charged silicon vacancies in semiconductor silicon carbide ($V_{Si}^{-}:SiC$) *(4)* and Neodymium rare earth ions embedded in the superconducting Niobium/Niobium-oxide film ($Nd^{3+}$:Nb-$NbO_x$) *(5)* that form the memory- provides access to the unique spin Hamiltonian that considers electronic and nuclear interactions. In this manuscript, we experimentally measure opto-electronic quantum-classical neurons integrated with a $V_{Si}^{-}:SiC$ optical spin qubit and examine how tailored neuronal spiking sequences drive the photoluminescence oscillations and modulate the resulting quantum spin transitions. We derive a quantum model for the system that considers the interaction between the neuron and optical qubit's spin components and perform calculations to project how quantum computing operations are expandable across the combined state space in the quantum-classical neuron and spin qubit (QN-SQ) system. For example, the resulting impact on the neuron's quantum states where the strength is adjustable due to dynamical memory changes. These results are critical for realizing hardware neurons with quantum optical capabilities and inspired by the existence in biology of biophotons that may participate in neural computing and networking processes.

## Introduction and motivation

The global research into hardware quantum-classical artificial neuron technology is accelerating and aims to fill a gap in capabilities for implementing the neuroscience known neuronal operations and for implementing proper biological-inspired quantum-integrated neuromorphic computing systems *(1-9)*. Achieving this also provides a needed upgrade in total computational capability and energy-efficiency over conventional approaches. Moreover, including optical processes improves the performance and provides the ability to network hardware neurons at larger scales for implementing advanced algorithms with physical neuronal qubit states with quantum optical spin Hamiltonian parameters. It has also emerged a desire to go beyond what's currently possible and unleash properties of biologically inspired behavior that are richer and more complex than what's possible with simple neural network models, for thus enabling systems that operate with concepts of decision making, emotions and even creativity *(10-14)*. Motivated by the activity of biophotons in biological systems *(15)*, we upgrade and network now a hardware neuron both with optically active rare earth ions embedded in the active region of the memory and optical spin qubits formed from silicon vacancy states in semiconductor silicon carbide to serve as optical spin qubits and perform direct measurements of the artificial neuron directing the modulation of the spin transitions and how this system performs optical computing.

We now characterize and examine a hardware neuron that consists of a optically gated memory formed from a optically transparent stack of indium-tin-oxide (ITO) and hafnium oxide ($HfO_x$) containing rare-earth ion doped Niobium oxide-Niobium enabling opto-electronic operation and the quantum-classical hardware neuron to accept and output optical and radio-frequency signals *(16-22)* that direct the quantum spin transition resonances of optical spin qubits formed from the silicon vacancy states *(23-29)* in semiconductor Silicon Carbide ($V_{Si}^{-}: SiC$). We examined the impact of a range of neuron-generated spiking sequences controlled optically through effectively varying the level of memory non-volatility due to the excitation of the ionic species and monitored through the resulting photoluminescence with near infrared 785.3 nm laser excitation. The ability to drive the optical qubit states is enabled through appropriately upconverting and modulating the burst spiking sequences and sweeping the optically detected magnetic resonance (ODMR) response under a wide band of 40.2 to 215.4 MHz and an applied magnetic field of up to 0.51 Tesla suitable for influencing the electron and nuclear hyperfine interactions and the resulting quantum oscillations. From the recorded experimental results, we then developed a quantum model for the system and examined in further detail the calculated response when including the spin Hamiltonian that considers the radiofrequency/optical excitation and the electron/nuclear spin interactions for spin $\frac{3}{2}$ with appropriate Pauli spin matrices. Finally, we show an example describing how distributed quantum operations can occur across this QN-SQ system, where the integration of the original hardware neuron and the optical spin qubits expand the state space for computing information for scenarios where the interactions are continuous to conditional. These results therefore provide important steps in advancing quantum-classical neuron technologies with quantum optical capabilities that are inspired by the elegance of the biological brain and the existence of biophotons that may participate in information processing.

## Design of quantum-classical hardware neurons interfaced to optical spin qubits

Figure 1(A) describes the architecture examined with the goal of interfacing quantum-classical hardware neurons (QN) that exhibit biologically inspired adaptive-itinerant spiking with optical spin qubits (SQ) to advance future computing. Central to this, we experimentally upgrade in our laboratory, a quantum-classical neuron implemented as described in *(1-2)* with rare earth $Nd^{3+}$ ions in the active memory region *(30-31)* to provide capability to directly optically modulate the degree of nonvolatility/volatility of the memristive $HfO_x$ film *(32-33)* of the superconductor-ionic quantum memory device *(2,3)*, and to provide an optical and microwave/RF interface for quantum memory operations that provide *(34-35)* necessary neuron-modulated signals that are compatible with optical spin qubits. These neuron-modulated signals are then directed along with synchronized optical pulses to drive and modulate spin transition resonances with the net result of modifying the quantum trajectories and providing access to the spin Hamiltonian to perform further advanced functions. A circuit schematic of the experiment performed is shown in Figure 1(B) and is adapted from the starting point in *(2)* where in this case the optical interface is shown and enabled with the integration of $Nd^{3+}$ rare earth ions and a memristor *(32,36)* that serves as a capacitively coupled interface to the semiconductor spin qubits after conversion of the signal to higher frequencies in the kilohertz to megahertz. We note that this circuit diagram is potentially analogous to the operation of biological neurons with an educated assumption that biophotons may participate in the functionality either through reactive oxygen species or from the action of microtubule elements. Understanding the role of optical photons in biological systems including brain action is a timely topic and for synthetic hardware development there are benefits to the integration of optical processes *(37-38)* to advance computing and networking operations.

The process for building the optically active memory devices integrated into hardware neurons is described in Figure 2(A-D) and begins with four-inch Silicon (Si) wafers with 300 nm of thermally grown silicon oxide ($SiO_2$). 90 nm of Niobium are deposited on the surface with DC sputtering at 300 Watts using a Nb target as the source. Electron beam lithography is performed with a PMMA e-beam resist and with calibrated dose and energy to produce mask defined features suitable for tunneling transport across. We found that a few nanometers of native oxide $NbO_x$ coat the surface and plasma dry etching with $SF_6$ chemistry is performed to define the structures. For the case of three-terminal gating experiments, 4.2 nm of Hafnium oxide are deposited by ALD i.e. 40-45 cycles followed by RF sputtering at 200 Watts of indium-tin-oxide to serve as an opto-electronic transparent gate. A second mask was utilized to define regions for implantation of rare-earth Neodymium ions, and the implantation process was performed and guided from simulations with the SRIM tool for selecting a proper dose and energy nominally 10 keV and $10^{12}$ $cm^{-2}$ in these experiments. The final memory device is shown in Figure 2(D) in both schematic formats that describes how the device is biased electrically and for applying radiofrequency (RF) and optical stimulus. Figure 2(E-F) are nominal images with probe tips probing a device as well as showing a fiber probe with optical laser light for excitation under both weak and strong intensity conditions. Figure 2(G), shows a cross-sectional schematic of the active gating region. Next, the process for building the optical spin qubits, i.e. from negatively charged vacancy states in semiconductor silicon carbide (SiC) is described in Figure 3(A-D) and begins with four inch 4H-SiC wafers that are subsequently diced to size and undergo thorough cleaning with acetone/isopropanol/deionized water. After that, ion implantations are performed of Carbon, i.e. $^{12}C$, where the dose and energy are guided from simulations and for these experiments a 20 keV and $10^{12}$ $cm^{-2}$ dose was found to be optimal. After implantation, the chips undergo rapid thermal annealing at 800-900 Celsius to form a large concentration of negatively charged vacancy states that behave as ensembles of optical spin qubits i.e. exhibit strong zero phonon lines (ZPLs). In this process, 2.1 nm of silicon oxide are deposited to passivate and protect the surface with atomic layer deposition and optically transparent electrodes are formed from RF sputtering of ITO. The ITO is defined by a lift-off process using conventional photolithography. A cross-sectional schematic of the active region of an optical spin qubit is shown in Figure 3(D,E) and a depiction of the crystalline arrangement with the atomic stacking showing a representative negatively charged vacancy state and an illustration of how both optical laser light and microwave neural oscillation pulses can serve as stimulus signals to modulate the optical spin transitions with optical micrographs shown in Figure 3(F-H).

## Measurements of hardware neurons modulating optical spin qubit resonances

Optoelectronic characterizations were performed by installing the $Nd^{3+}$:$NbO_x$-Nb and $V_{Si}^-$:SiC chips inside the cryo-magneto-optical probe station together with Nb-$HfO_x$-Nb devices. Electrical inputs were provided by direct probing with Tungsten probe tips. Radiofrequency (RF) signals were applied via a probe tip attached with a coil antenna and fiber optic cables with exposed probes were utilized to provide optical laser illumination. Optical emission was collected by focusing the lens of the zoom microscope to a desired active region with numerical aperture (NA=0.24) on a spot size of 12 microns, obtaining feedback from images using a CCD camera from a beam split portion of the optical signal. The optical emission was collected into an optical fiber and routed for detection with a VIS/NIR spectrometer with a cryogenically cooled camera to detect the emission across a wide 700 to 1200 nm range. For optical excitation with a 785.3 nm laser, an 830 nm long pass filter is used to detect the emission. The laser for optical excitation is pulsed appropriately and synchronized to the neural spiking signal where a set time is introduced between RF and optical pulses to enter quantum optical regimes of operation. Burst spiking signals from neural circuits are obtained from appropriate biasing of Nb-$HfO_x$-Nb devices with integrated $Nd^{3+}$:$NbO_x$-Nb stacks, and where optical illumination provides for adjustment of the non-volatility *(32)* through electronic and optical modulation of the Oxygen and $Nd^{3+}$ ions.

Figure 4 presents characterization of the optical memory formed from Nb-$HfO_x$-Nb with the ITO gating and integrated with $Nd^{3+,}$ i.e. the optically active $Nd^{3+}$:$NbO_x$-Nb film stacks. Representative chips are characterized first for their memory/memrisitve properties at room temperature and ability to optically modulate. The film has a crystal structure based on a body-centered-cubic (BCC) configuration with 2-3 nm of native $NbO_x$ as depicted in Figure (4A) and the experimental EDX measurements in Figure 4(B) confirm Nb with a 1-2% concentration of Nd. For details on these films' material aspect properties, please see our previous work in reference *(30)*. Figure 4(C) and Video S1 shows an optical micrograph while under combined electrical, microwave, and optical biasing and stimulus and Figure 4(D) shows a significant bright spot where strong emission is observed and where our zoom microscope is focused to collect the emission that represents ensembles of highly luminescent $Nd^{3+}$ centers where in the spectroscopic evaluation the optical power is reduced appropriately to the necessary level. Three terminal current-voltage (IV) measurements with clear indications of optical modulation of the nonvolatility and switching mechanisms of a device measured with and without the light are shown are in Figure 4(E), where the laser power density is 8 mW/$\mu m^2$ at 785.3 nm. Figure 4(F) are measurements of the optical emission spectrum across the 820-1140 nm wavelength range at room temperature and in the superconducting regime of operation i.e. at 8.4 Kelvin under conditions of increasing magnetic and electric fields that update the transitions probabilities of the primary $4F_{3/2}$ -> $4I_{9/2}$ transition and Figure 4(G,H), a depiction of these levels *(19)* including also the $4F_{3/2}$ ->$4I_{11/2}$. Next, direct measurements are done of a neural circuit as shown in Figure 4(I) with integration in the feedback oscillator of the devices demonstrating capability to adjust the spiking. This was done through biasing of the circuit and highlights the cases of mixed positive and negative feedback modes as well as just strong negative feedback; where true burst modes are isolated. Figure 4(J) shows an example of measurements done with and without optical stimulus demonstrating changes in the spiking modes due to alteration of the films nonvolatility through excitation of the $Nd^{3+}$ along with Oxygen ionic complexes as described in the energy level diagram.

Regarding the optical spin qubits formed from $V_{Si}^-$:SiC, i.e. negatively charged silicon vacancy states in SiC, where the vacancy surrounded $sp^3$ dangling bonds capture an extra electron, thus totaling five electrons, we describe here in brief the critical quantum features relevant to operation of this hardware neuron-spin qubit system which have good properties in behaving as photon emitters and performing quantum sensing because they have long spin coherence times and optical transitions *(20-24)* where emitted photons can be entangled with spin states. Figure 5(A) shows a respective energy level manifold that describes the key features of the operation highlighting the impacts of Zeeman splitting from applied AC/magnetic fields as well as further Stark splitting from applied electric fields with appropriate alignment with respect to the wafer's c-axis crystal orientation providing a wide array of spin transitions. Optical excitation and emission measurements are done on sites discovered to behave as high quality optical spin qubits on spots such as shown in Figure 5(B) to confirm the presence of high-quality spin transitions as described in Figure 5(C)/Table I where there are priority main transitions, including for example $m_s = |\pm 3/2_{GS}\rangle \leftrightarrow |\pm 3/2_{ES}\rangle$ with $\Delta m_s = 0$ as well as opening of spin mixing possibilities with especially the application of applied fields through isolating of the S=3/2 type defect qubit states and observance of their zero-phonon-lines (ZPL) at room and cryogenic temperatures, shown in Figure 5(D), isolating the ZPL at a baseline temperature of 8.4 Kelvin. Lifetime measurements are shown in Figure 5(E) and performed by pulsing the laser and exciting the spin qubits in a micrometer sized spot when brought near the active region. Figure 5(F) presents a collection of emission spectra impacted by the neuron-modulated RF spiking signals across the 75.4 to 120.1 MHz range and for conditions of the application of an electric field. Shown specifically are cases where strong resonance actions occur as that can be connected readily to calculations of the optically detected magnetic resonance responses that guide the experimental measurements. Lastly, for experimental quantum photonic device measurements Figure 6(A) shows the complete arrangement including all aspects of the hardware neuron-optical spin qubit system. Figure 6(B) shows an example of a neuron-modulated spiking signal where optical excitation of the $Nd^{3+}$ rare earth ions uniquely adjusts the spiking mode and Figure 6(C) and Videos S2-S3 shows the photoluminescence oscillations. A further example is shown in Figure 6(D) after a fixed time and the final PL oscillations measured directly from the $V_{Si}^{-}$:SiC are presented in Figure 6(E) and Figures (S1,S2).

## Upgraded model of the quantum neuron coupled to the spin qubit (QN-SQ)

To examine the ability for performing advanced computations with this system, we analyze the response theoretically through modeling and simulations. The combined total Hamiltonian for our experimental system consists of the components from the spin qubit (SQ) formed from negatively charged vacancy states in semiconductor SiC, and the quantum neuron (QN) states i.e. the quantum neuron with superconductor-ionic memory- in this case embedded with rare earth ions, and now advanced in this research with the complex interaction between the quantum neuron and the optical spin qubit where the spin qubit Hamiltonian expressed as:

$$H_{SQ}(t) = D(S_z^2 - \tfrac{5}{4}) + \mathrm{d}E(S_x^2 - S_y^2) + \mu_B \vec{B}(\omega t) \cdot g \cdot \vec{S} + \vec{S} \cdot A \cdot \vec{I} - \gamma_n \hbar \vec{B} \cdot \vec{I} - H_{applied_E}(t) \quad (1)$$

Where in this work, the applied electric field perturbation as $H_{applied_E}(t) = d_z F_z(t) S_z^2$ (2), where $E$ is the rhombic splitting parameter caused by local strain or crystal asymmetry. $F_z$ is the strength of the externally applied electric field directed along the z-axis. $D$ is the longitudinal Zero-Field Splitting (ZFS) parameter for V1, $\gamma_n$, is the nuclear Gyromagnetic Ratio, g is the electron g-Tensor and $d_z$ is the electric field dipole coupling constant. The 4x4 spin matrices for S=3/2 are:

$$S_x^{(3/2)} = \frac{\hbar}{2}\begin{pmatrix} 0 & \sqrt{3} & 0 & 0 \\ \sqrt{3} & 0 & 2 & 0 \\ 0 & 2 & 0 & \sqrt{3} \\ 0 & 0 & \sqrt{3} & 0 \end{pmatrix} \quad (3) \qquad S_y^{(3/2)} = \frac{\hbar}{2}\begin{pmatrix} 0 & -i\sqrt{3} & 0 & 0 \\ i\sqrt{3} & 0 & -2i & 0 \\ 0 & 2i & 0 & -i\sqrt{3} \\ 0 & 0 & i\sqrt{3} & 0 \end{pmatrix} \quad (4)$$

$$S_z^{(3/2)} = \frac{\hbar}{2}\begin{pmatrix} 3 & 0 & 0 & 0 \\ 0 & 1 & 0 & 0 \\ 0 & 0 & -1 & 0 \\ 0 & 0 & 0 & -3 \end{pmatrix} \quad (5)$$

and the time dependent coupling with coefficients $g_{coup,k}$:

$$H_{int}^{QN-SQ}(t) = \sum_{k=1}^{N}\left(g_{coup,k}(t)\left(\sigma_{ge}^{QN} + \sigma_{ge}^{QN\dagger}\right)\otimes S_x\right) \quad (6)$$

The quantum neuron Hamiltonian with time shift $d$, where $\sigma_{ge}$ are the GS to ES transitions is:

$$H_{QN}(t,\theta) = \sum_{k=1}^{N}\left(-g_k sin\left(\frac{\theta_k}{2}\right)\left(\sigma_{ge}^{\dagger}a + a^{\dagger}\sigma_{ge}\right) - Acos(\frac{\theta_k}{2})(\sigma_{ge}^{\dagger} + \sigma_{ge})\sin\left(\left(\frac{t}{e}\right)^2 - d\right)\right) \quad (7)$$

Therefore, the total combined and now upgraded Hamiltonian for the designed QN-SQ system is:

$$H_{total}^{QN-SQ}(t,\theta) = H_{QN} + H_{SQ} + H_{int}^{QN-SQ} \quad (8)$$

## Quantum information operations directed by neuron-spin qubit interactions

Access of the system to the spin Hamiltonian and spin operator's degree of freedom, makes it possible to implement quantum operations that perform a distributed quantum information processing making use of the coupling interactions in the composite system. To analyze this, we performed dynamical time dependent simulations with the quantum master equation approach employing a tensor transfer technique and accounting for non-Markovianity. We examined the cases of (i) continuous coupling interactions $\sigma_{ge}\hat{S}_y$ and (ii) conditional interactions $\sigma_{ge}(a\hat{S}_y + b)$ where the dynamical ionic memory impacts qubit neuron learning through the *a* and *b* parameters. Figure 7(A-E) shows calculated ODMR spectra with and without the application of neuron-modulated spiking signals that introduce an electric field effect and where the transition probabilities are computed. These calculations provide points for conditions that produce good resonance supporting the experiments and quantum information operations described now. Figure 8(A-D) shows simulation results with and without the optical spin qubit upgrade and under pulsed conditions that mimic the experiments with neuron-modulated spiking, effectively contributing to a function that's analogous to π-π/2 Rabi sequences *(39-40)*. Figure 8(E,F) shows the extracted coherence for the quantum neurons and optical spin qubits that track a millisecond duration and the entanglement concurrence metrics for the full QN-SQ system. The function approximates a spiking signal expressed where the constant *d* introduces time shift. Results for several effective pulse widths are shown with simulation outputs across the Bloch Sphere are shown and the method was calibrated to produce a clear quantum information operation and confirmed by analyzing the coherence levels of both the qubit-neuron state and the spin state of the optical qubit. Practically, a large-scale network of these hardware neurons with integrated optically active superconductor-ionic memories and optical spin qubits is capable of performing brain inspired dynamical computational operations as described in Table II and Figure 9 that summarizes the key components and operational modalities of the upgraded quantum-classical neurons.

## Conclusion

We upgraded hardware neurons by incorporating optically active rare-earth-ions into the memory and through interfacing to optical spin qubits formed from negatively charged vacancy states in semiconductor SiC. This provides the system with advanced computational capabilities inspired by the potential role of biophotons in biological information processing and through the neuron-modulated electronic and magnetic coupling interactions of the spin Hamiltonian in conjunction with the spiking processes and dynamics. Our architecture includes optically triggered neuron spiking processes that modulate the spin transition resonances of the optical qubits and where experimental adjustment of the timing and synchronization with laser excitation results in Bloch sphere rotations where the photoluminescence oscillations from the optical qubits are directly observable. We examined how this quantum neuron-spin qubit (QN-SQ) system performs advanced computing processes through simple interactions designed from continuous to conditional where the impacts from the coherence and entanglement concurrence are observable. These results are thus of critical significance for advancing biological brain-inspired computing.

**Supplementary Materials:**

**The PDF file includes:**

Materials and Methods

Supplementary Figures

Figs. S1-S2

**Auxiliary Supplementary Materials for this manuscript include the following:**

Videos S1 to S3

**Acknowledgments:**

We're grateful for the NIWC Pacific information technology and laboratory infrastructure. This research was conducted at the NIWC Pacific quantum (memory) lab. OMNs use of the Qualcomm Nano[3] facilities at the University of California, San Diego is also acknowledged.

**Funding:** This work was supported by the Naval Information Warfare Center (NIWC) Pacific. Distribution Statement A. Approved for public release: distribution is unlimited.

**Competing interests:** OMN declares issued US Patents (1) "Quantum computing with photonic/ionic tuning of entanglement", US Patent US10,133,986B1 (2) "Formation of field-tunable silicon carbide defect qubits with optically transparent electrodes and silicon oxide surface passivation", US Patent US10262871B1 and patents pending (3) "Multifunctional Quantum Node Device and Methods", US Patent Application US20210020821A1

**Data and materials availability:** All data are available in the main text or supplementary materials" Several videos are included in the auxiliary supplementary materials. Requests for additional data or questions should be made to the corresponding author.

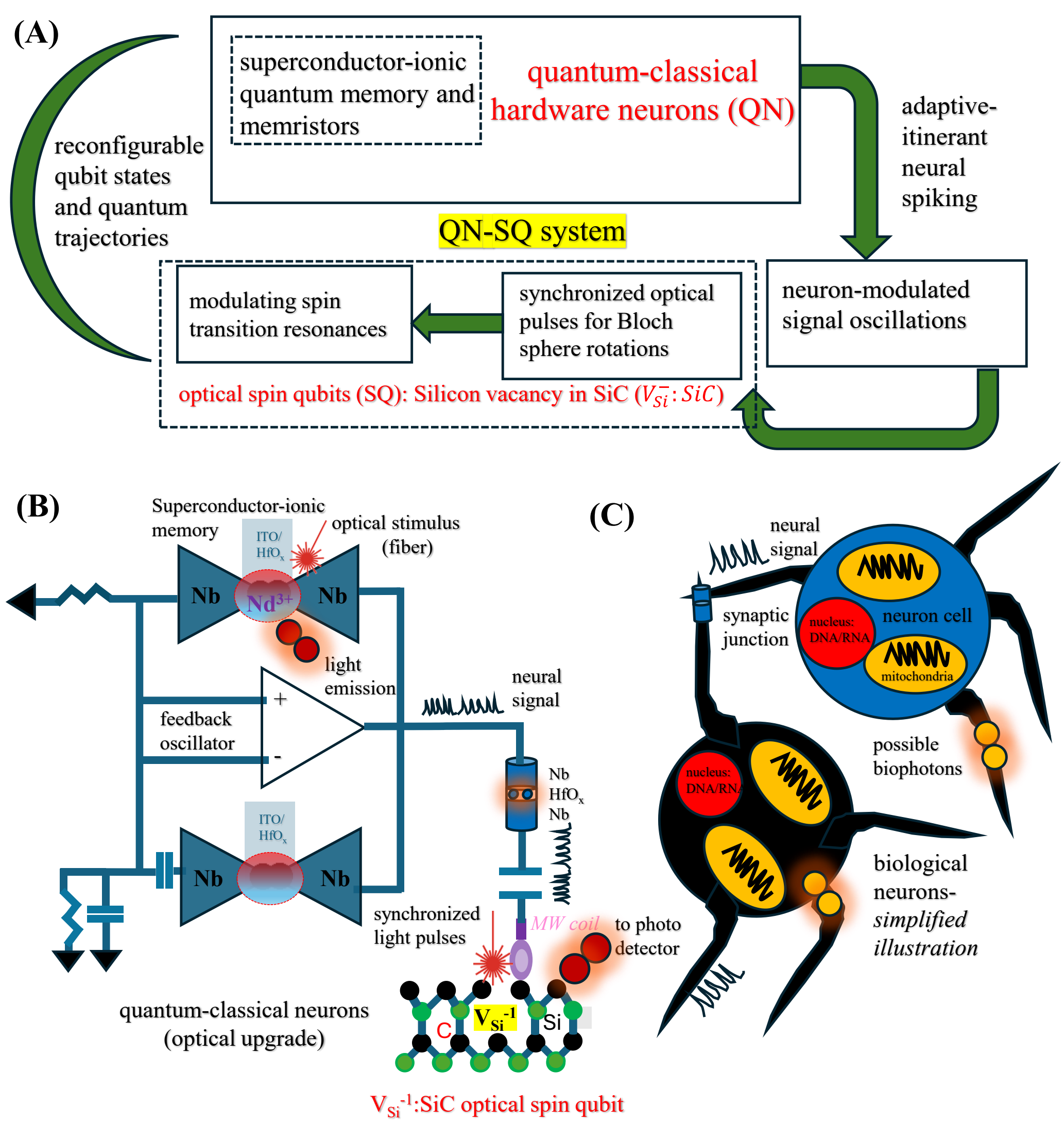

**Figure 1: Quantum-classical neuron and optical spin qubit (QN-SQ) architecture: (A)** Primary features highlighted include the incorporation of optically active rare-earth ions into the superconductor-ionic quantum memory and the use of neural spiking sequences in conjunction with optical pulses to modulate the spin transitions of the silicon vacancies in semiconductor SiC, serving as optical spin qubits in this work. **(B)** Circuit diagram showing all the opto-electronic components and annotated to describe the propagation of the routed neural oscillations and synchronized optical excitation stimulus signals that are coupled to and received by the optical spin qubits in order to perform advanced computing and networking quantum information operations. **(C)** Side-by-side comparison with a simplified biological neuron illustration, depicting some of the key features and the possible role of biophotons that may participate in neuronal information processing mechanisms- thus a motivating factor for integrating optical spin qubits into our quantum-classical hardware neuron technology *(1-3)*.

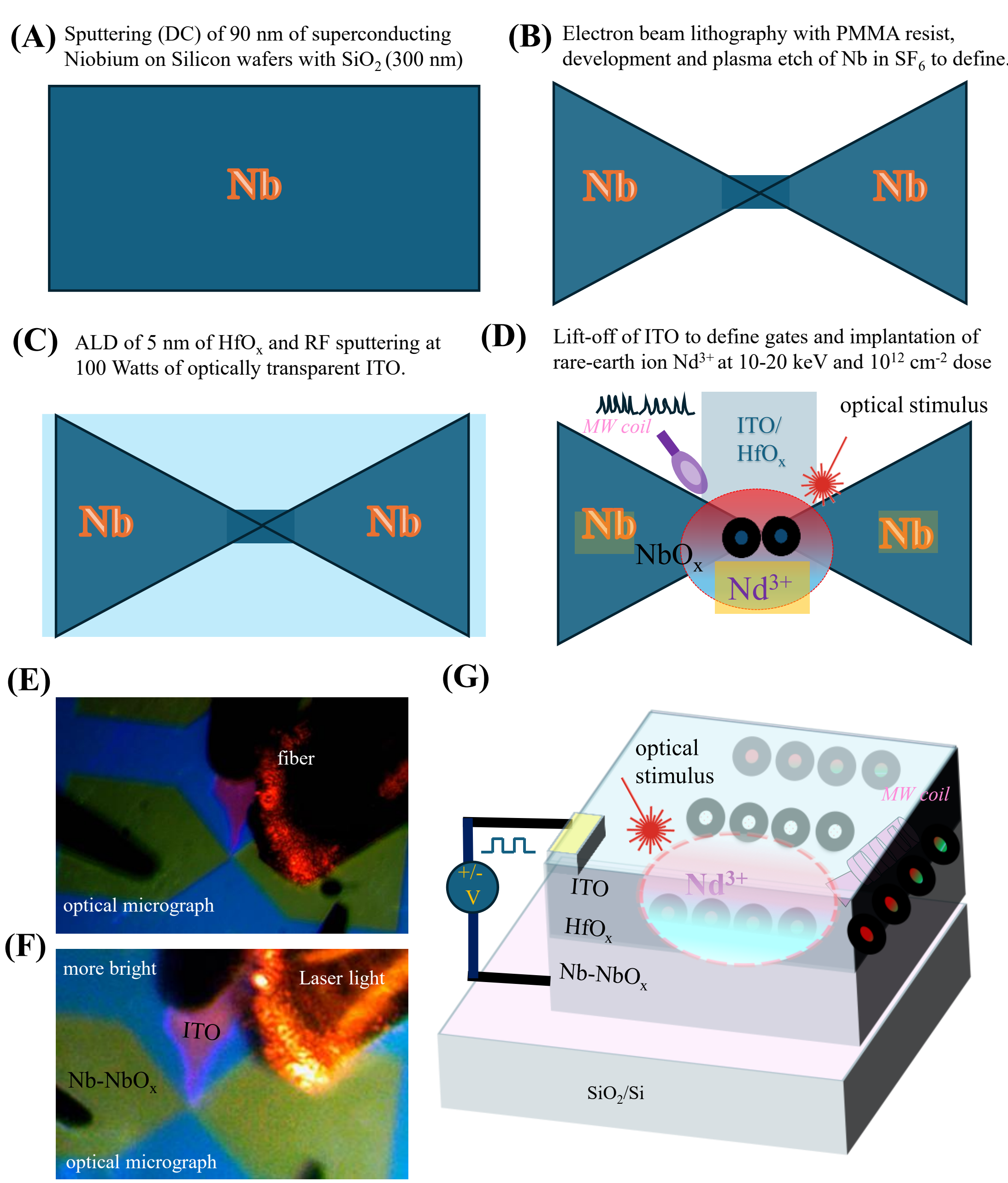

**Figure 2: Process flow for rare-earth ion integrated $Nd^{3+}$:Nb-$NbO_x$ memory (A)** Thin film deposition. **(B)** Electron beam lithography and etch. **(C)** ALD and sputtering of optically transparent dielectrics and **(D)** implantation of Nd at specific dose and energy to form luminescent $Nd^{3+}$ complexes and final lift-off to define gates. **(E)** Top-down view optical image of device under test showing illumination of the active region with a fiber optic probe **(F)** with increased intensity illumination. **(G)** Cross-section of the measurement configuration.

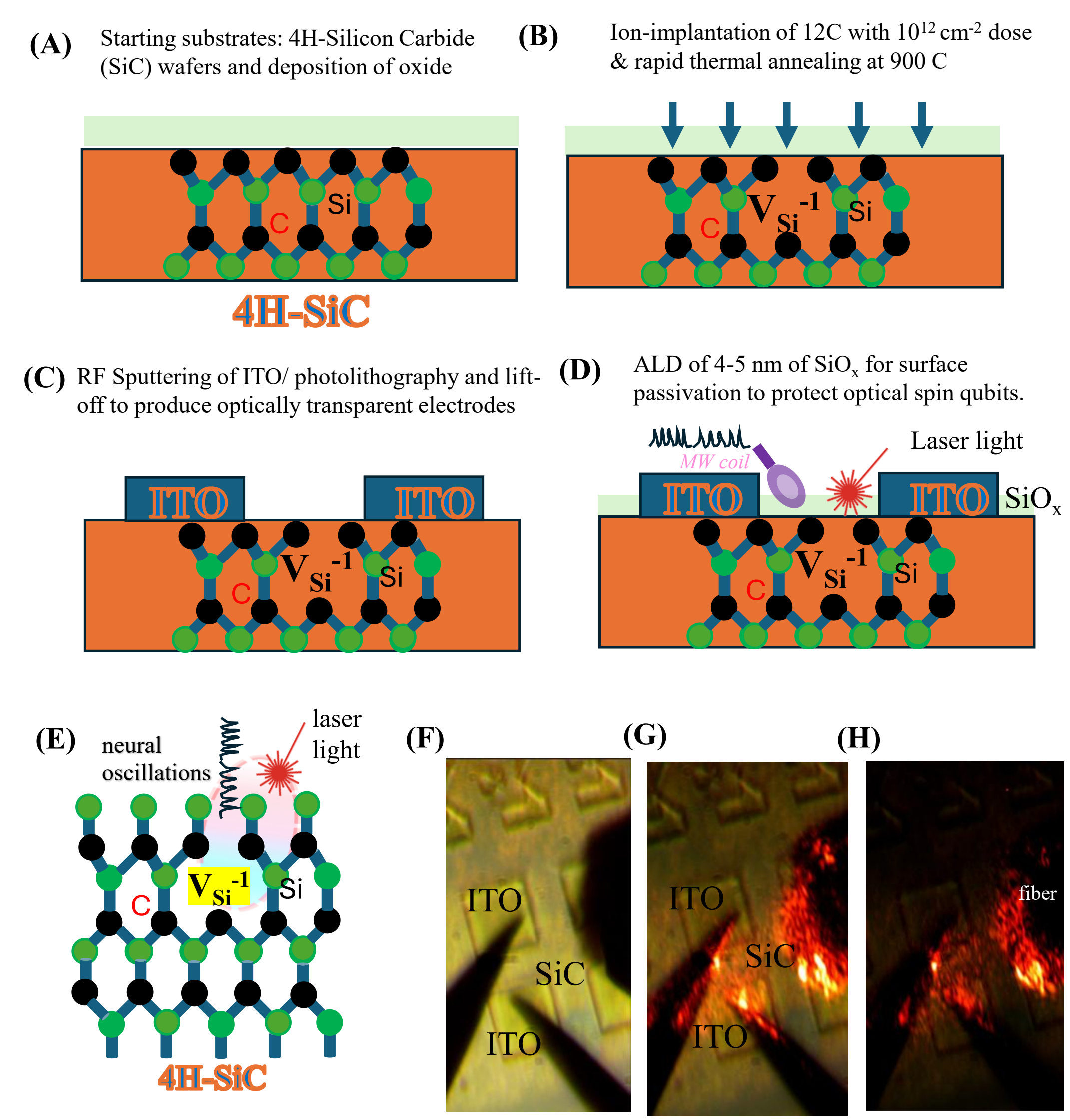

**Figure 3: Process flow for semiconductor $V_{Si}^-$:SiC optical spin qubits (A)** 4H-SiC starting substrates with $SiO_x$ interfacial layer and sufficient intrinsic doping density were initially diced and cleaned. **(B)** Ion implantation of Si and the subsequent high temperature rapid thermal annealing were performed to form a concentration of negatively charged Si vacancy states that behave as optical spin qubits with pronounced zero-phonon-line emission. **(C)** RF sputtering of optically transparent ITO electrodes and photolithography with a lift-off process. **(D)** Subsequent ALD deposition of a few nanometers of $SiO_x$ that serves as surface passivation and to provide an opto-electronic interface to a completed device in **(E)** for applying **(F-H)** DC voltage/electric field while retaining the capability for performing measurements of the optical emission while interacting with stimulus signals generated from integrated neurons.

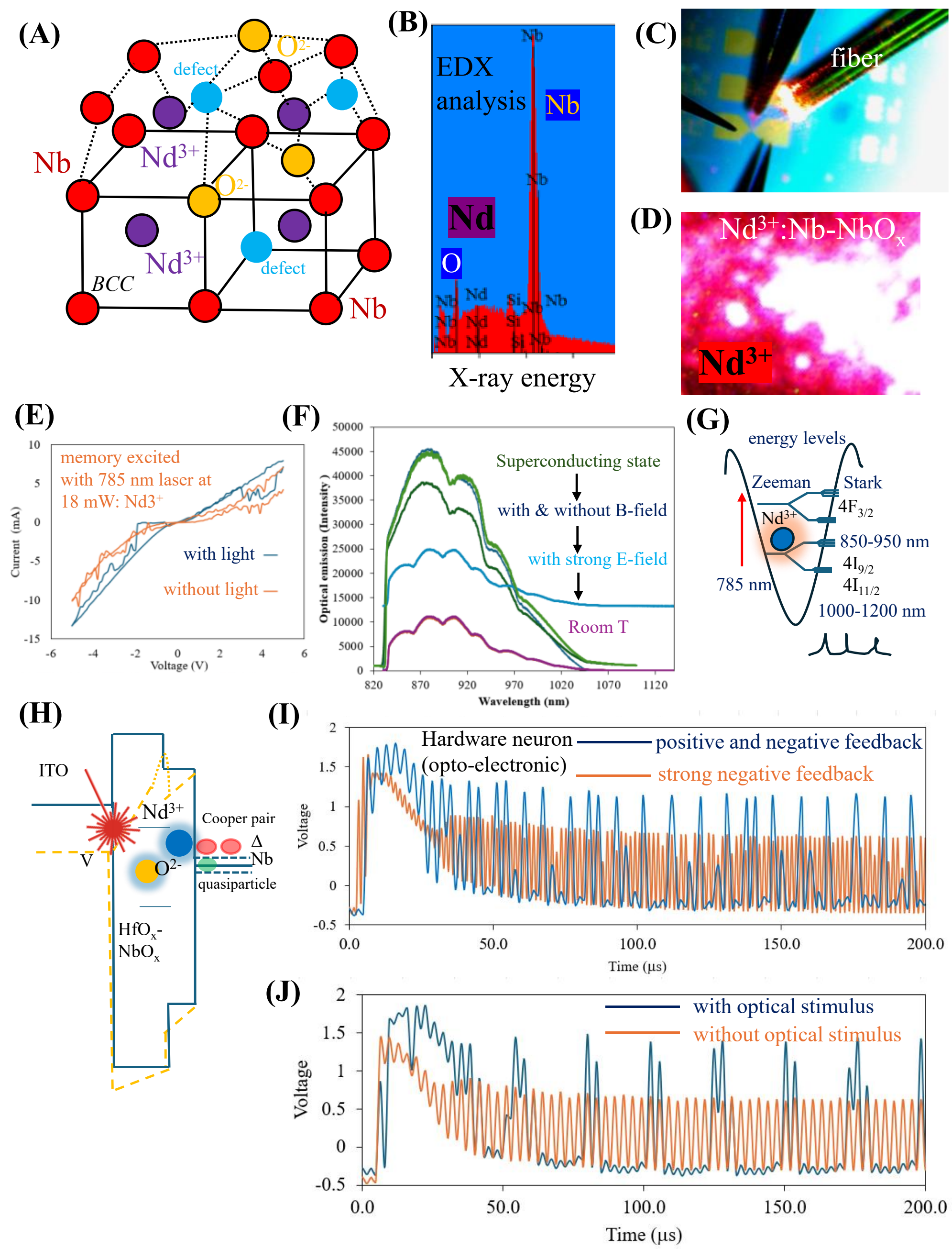

**Figure 4: Optoelectronic memory and spiking neuron (A)** Crystallographic diagram. **(B)** EDX spectra from *(5)* confirming Nb, and Nd at 1-2 %. **(C)** Experimental measurement configuration under optical illumination. **(D)** A zoom-in micrograph. **(E)** Current-voltage characteristics with/without light. **(F)** Collected optical emission spectra across the NIR band at both the room temperature and superconducting states under magnetic and electric fields. **(G)** $Nd^{3+}$ energy levels and **(H)** energy band diagram. **(I, J)** Impacts on the neural spiking response with/without optical stimulus that effectively adjusts the non-nonvolatility strength.

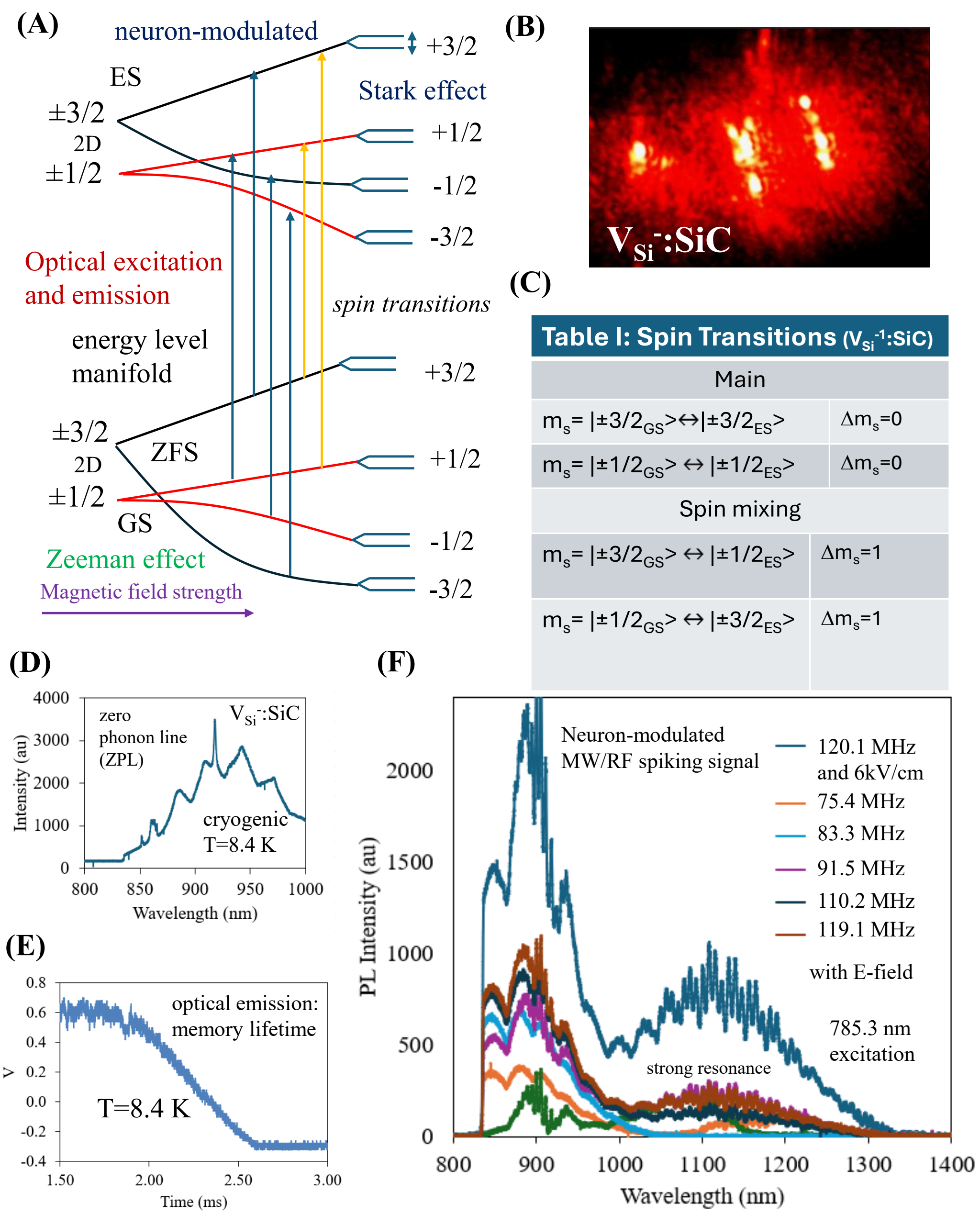

| Table I: Spin Transitions ($V_{Si}^{-1}$:SiC) | |
|---|---|
| Main | |
| $m_s$= $\lvert\pm3/2_{GS}\rangle \leftrightarrow \lvert\pm3/2_{ES}\rangle$ | $\Delta m_s$=0 |
| $m_s$= $\lvert\pm1/2_{GS}\rangle \leftrightarrow \lvert\pm1/2_{ES}\rangle$ | $\Delta m_s$=0 |
| Spin mixing | |
| $m_s$= $\lvert\pm3/2_{GS}\rangle \leftrightarrow \lvert\pm1/2_{ES}\rangle$ | $\Delta m_s$=1 |
| $m_s$= $\lvert\pm1/2_{GS}\rangle \leftrightarrow \lvert\pm3/2_{ES}\rangle$ | $\Delta m_s$=1 |

**Figure 5: Neuron modulation of negatively charged Si vacancy in SiC optical spin qubit** **(A)** Energy-level manifold (simplified) describing the spin transitions and how their neuron modulation can impact through the Zeeman and Stark effects. **(B)** Optical emission attributed to the strong ZPLs in the optical spin qubit system i.e. $V_{Si}^-$:SiC while inserted in a vacuum chamber that houses the superconductor-ionic memories, memristors, and optical spin qubits. **(C)** Table I- spin transitions with emission characteristics **(D)** at cryogenic temperatures and the **(E)** optical memory lifetime measured from an Avalanche photodiode. **(F)** Measured optical spectra for several cases of neuron-modulated signals with various RF and electric fields.

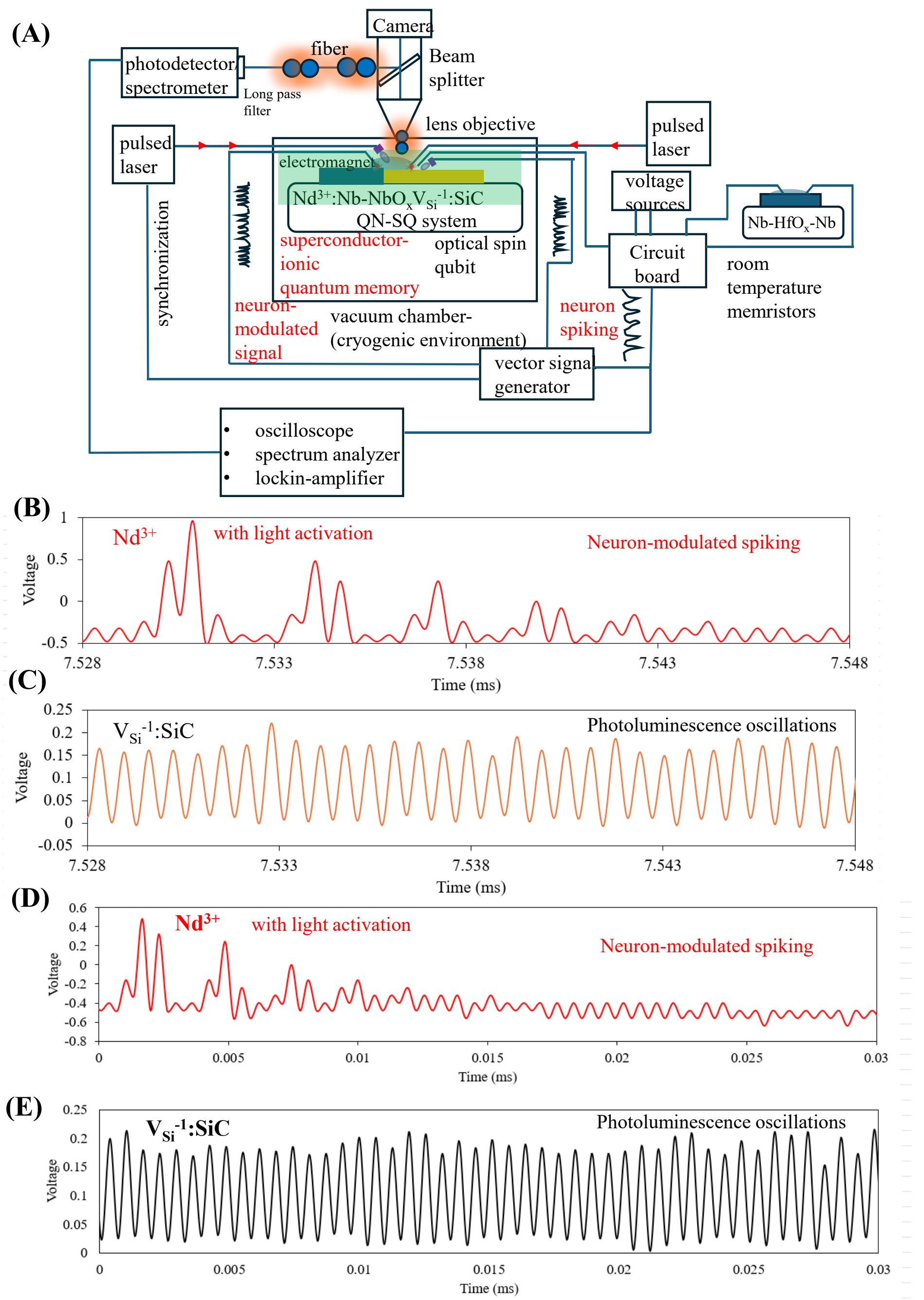

**Figure 6: Characterized quantum neuron-spin qubit (QN-SQ) full system** **(A)** Experimental measurement configuration. **(B)** Example output of the optically excited hardware neuron spiking sequence with a segment of the neuron-modulated RF pulses that are then synchronized to optical excitation pulses to excite the ensembles of $Vsi^-$:SiC optical spin qubits and the resulting **(C)** time dependent photoluminescence characteristics conducting quantum information operations. **(D-E)** The same type of quantum measurements performed, but after a definite and calibrated time i.e. to ensure sufficient rotation across the Bloch sphere.

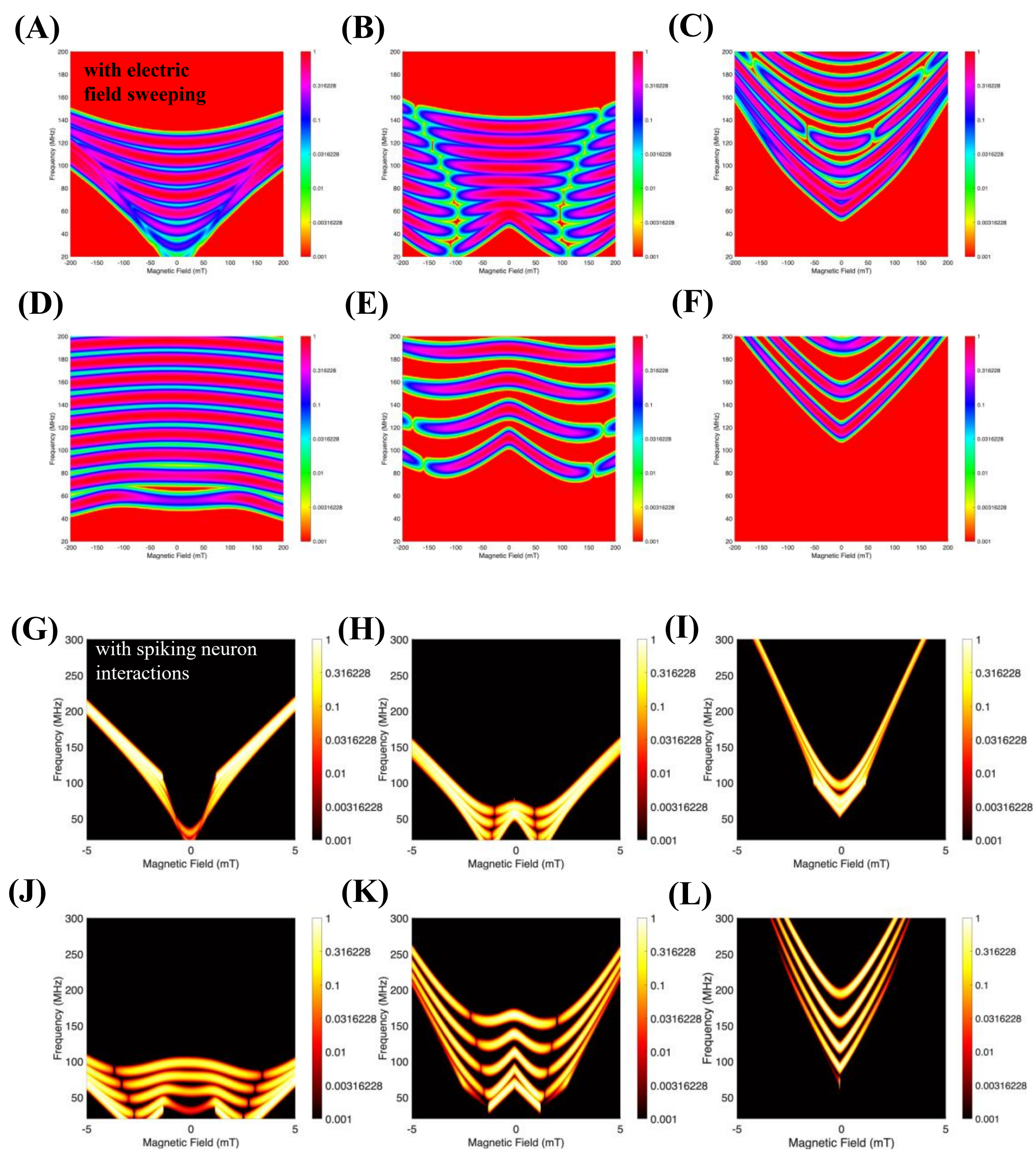

**Figure 7: Calculated optically detected magnetic resonance (ODMR)** of the main spin transitions for modest magnetic fields that guide and calibrate the measurement conditions for quantum optical experiments in order to observe resonant conditions and the emergence of quantum oscillations in the photoluminescence. **(A-F)** ODMR calculations up to 0.2 T for a $V_{Si}^{-1}$:SiC optical qubit system without coupling interactions and **(G-L)** when considering a strong coupling with the hardware neurons and also the direct Radiofrequency (50-300 MHz) modulation- taking account of the Stark effect- and an applied magnetic field up to up to 5 mT.

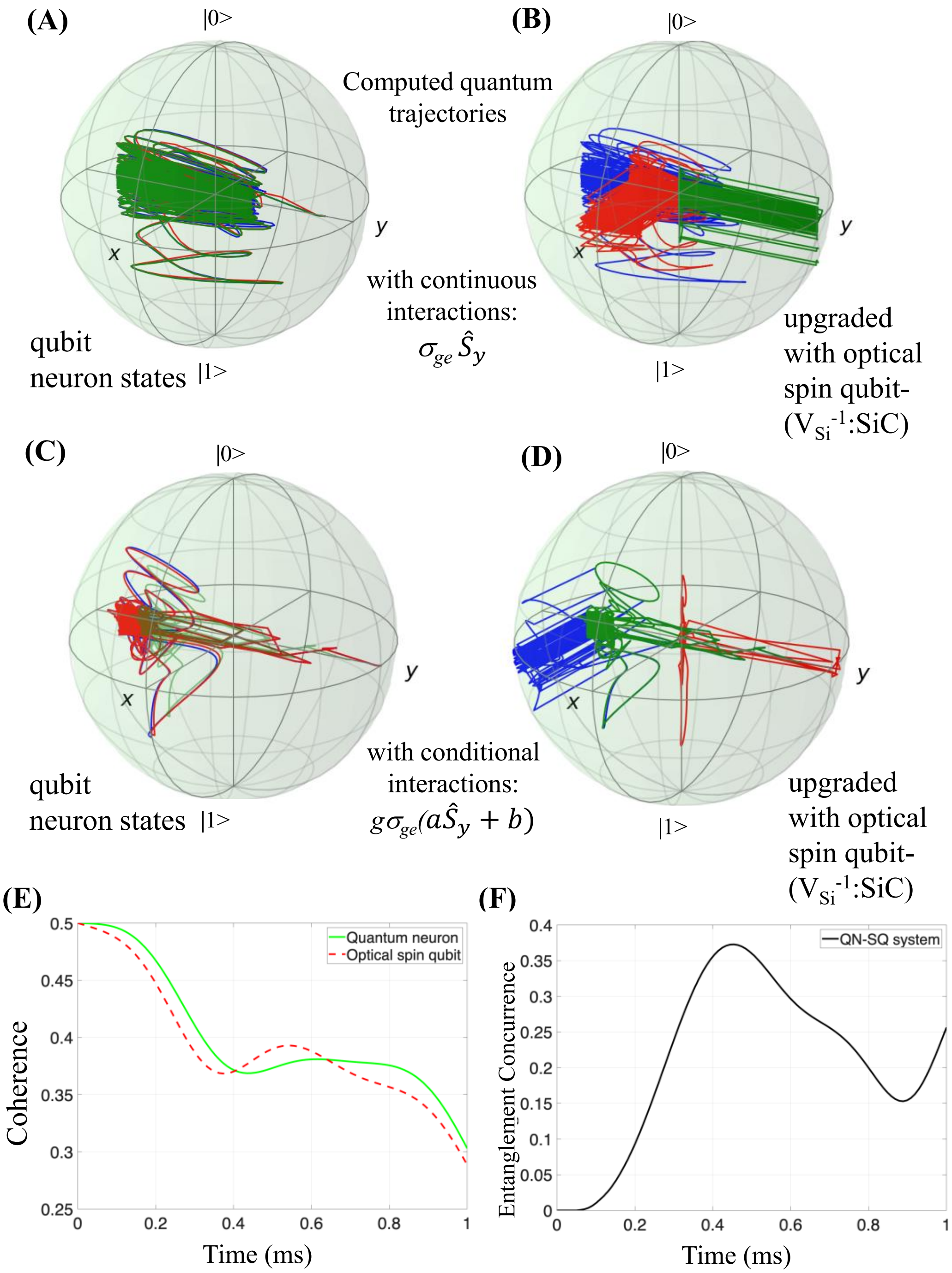

**Figure 8: Qubit neuron states** (left) and their calculated quantum trajectories with and without the optical spin qubit upgrade (right) for the cases of **(A-B)** independent qubit neurons and **(C-D)** conditionally coupled qubit neurons. **(E)** Calculated time-dependent quantum coherence metrics for the individual units and **(F)** the entanglement concurrence for the QN-SQ system.

| Table II: Quantum neuron-spin qubit (QN-SQ) system | | |
|---|---|---|
| Hardware components | Operational modalities | Composite measures |
| ➢ Quantum-classical neuron (QN) with superconductor-ionic (rare-earth $Nd^{3+}$ embedded) quantum memories that emit in the NIR and integrated room temperature memristors (1-2) | • Adaptive-itinerant spiking behavior<br>• Coherence/entanglement concurrence<br>• Learning/training algorithms<br>• Qubit states and quantum trajectories | • Neuron-directed modulation of spiking<br>• Quantum algorithms across the Bloch sphere<br>• Non-Markovianity and information generation<br>• Quantum oscillations and advanced Hamiltonians<br>• Quantum computing and networking operations |
| ➢ Optical spin qubits (SQ) formed from negatively charged silicon vacancies in semiconductor SiC ($V_{Si}^{-}$:SiC) that emit in the NIR and respond to MW stimulus (5-6) | • Optical qubit light emission<br>• Spin transitions rules and selection<br>• Optically-detected magnetic resonance<br>• Zeeman and Stark effects | |

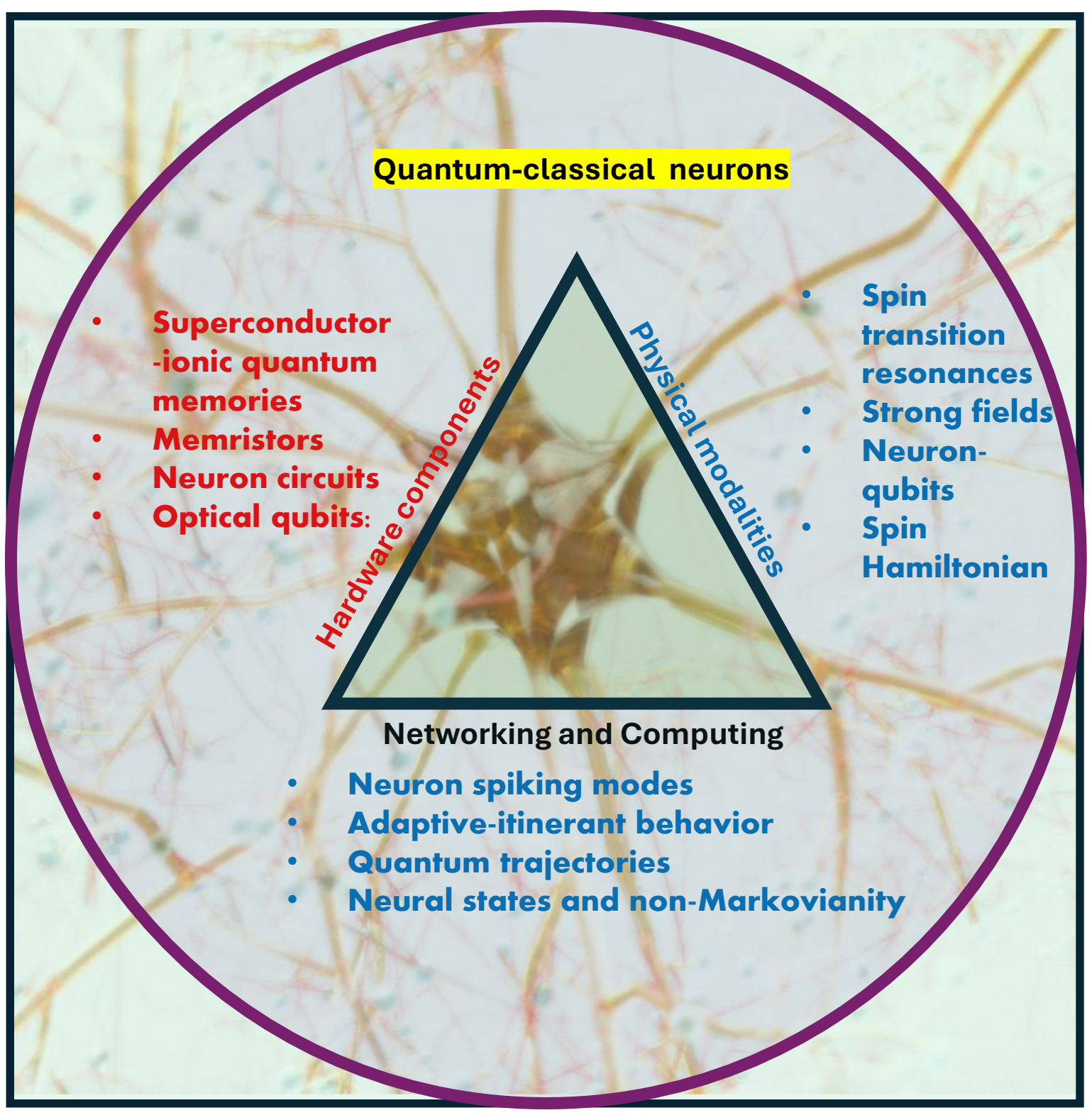

**Figure 9: Visual representation of the design space for quantum-classical neurons** highlighting in table II the major features advanced in this work including (1) technological components: superconductor-ionic memories and optical spin qubits (2) physical modalities related to the neuron-qubit states and optical spin transitions and (3) networking and computing functions that represent neuron spiking modes and the quantum trajectories of the neural states.

# Supplementary Materials for

## Quantum-classical neurons upgraded with optical spin qubits for advanced computing and networking architectures

**Authors:** Osama M. Nayfeh[1†]

**Affiliations:** [1] Basic and Applied Research Division, Code 71780, Cyber/Science and Technology Department, Naval Information Warfare Center (NIWC) Pacific, Naval Base Point Loma, San Diego, CA USA

†**Corresponding Author:** osama.m.nayfeh.civ@us.navy.mil; (619) 553-2770

**The PDF file includes:**

**Materials and Methods**

**Supplementary Figures**

Figs. S1-S2

**Auxiliary Supplementary Materials for this manuscript include the following:**

Videos S1 to S3

**Materials and Methods:**

The QN-SQ system was designed and experimentally measured where the superconductor-ionic quantum memories with rare-earth embedded ions and the silicon vacancy in SiC optical spin qubits installed together in the same vacuum chamber housed in a cryo-magneto-optical probe station (MicroXact Inc). For details on micro/nanofabrication processes see the descriptions in the main text and references *(4-5,30-32).* The superconductor-ionic memories were integrated directly into the feedback oscillator circuit as described in further detail in *(1-3)*. Calibration was performed to isolate some specific spiking sequences for demonstration, and the resulting hardware neuron spiking signal was an input to modulate a MHz pulsed signal. The resulting neuron-modulated RF signal was then inputted with a coil antenna and directed at the active region that contains optical spin qubits. The pulsed signal with an additional delay was utilized to also trigger a pulsed optical laser signal generated with a Thorlabs laser driver and a fiber optical butterfly quantum well laser diode. The combined optical/RF signals served as inputs to modulate the ODMR response of the optical spin qubits and the dependent quantum oscillations in the photo luminescence were collected. To perform measurements of the optical emission- fiber optic, DC, and RF probes were operated with a zoom microscope integrated with a spectrometer and cryogenically cooled camera.

**Supplementary Figures**

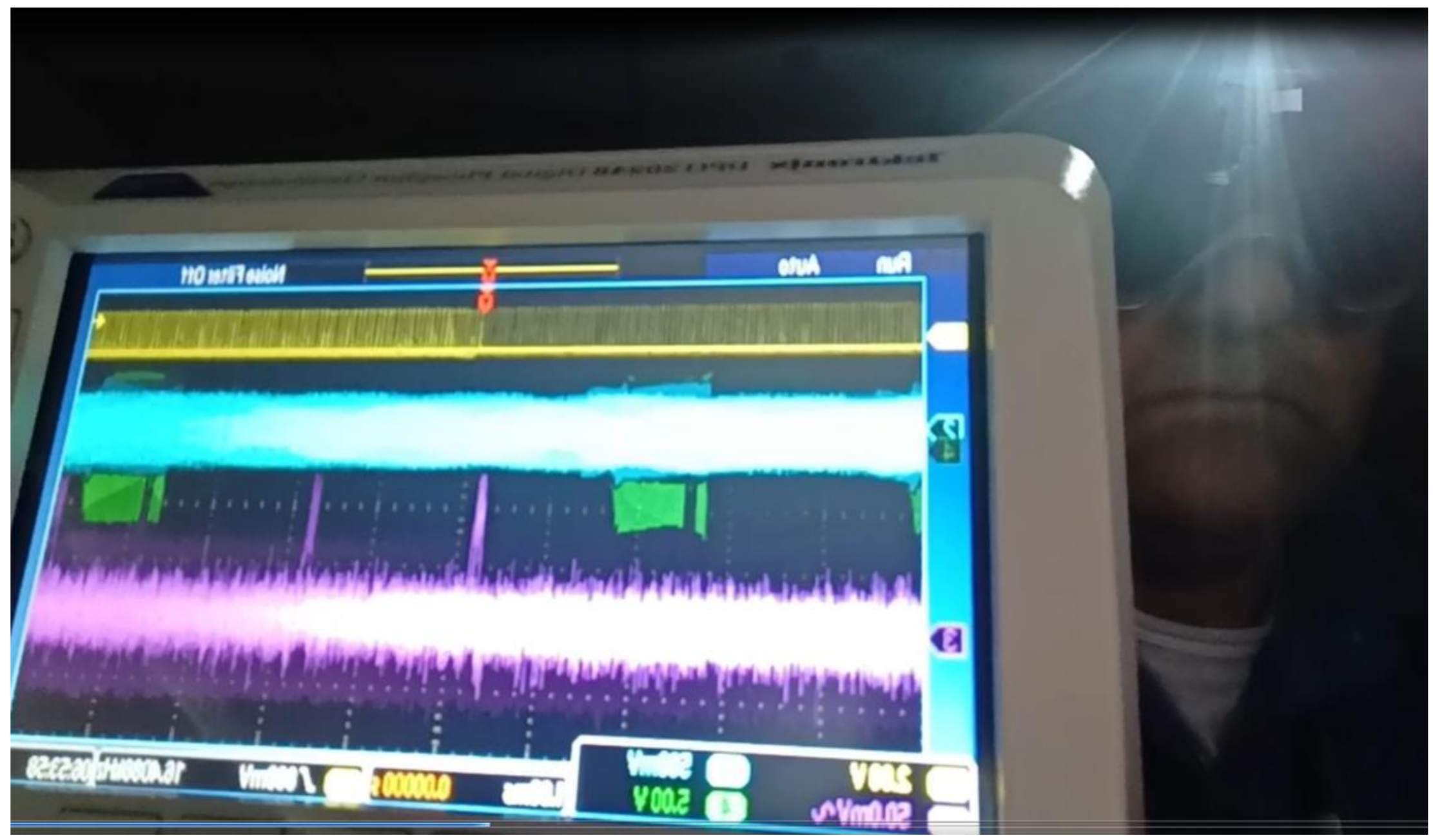

**Figure S1:** Example of quantum information operations performed with our QN-SQ system in the laboratory. The neuron spiking sequence is shown in yellow, and the green/turquoise are the neuron-modulated RF signals. The magenta shows the measured photoluminescence (PL) signals where the timing sequence is synchronized and calibrated to produce PL oscillations in the optical spin qubit’s emission due to neuron-modulation of the spin transitions.

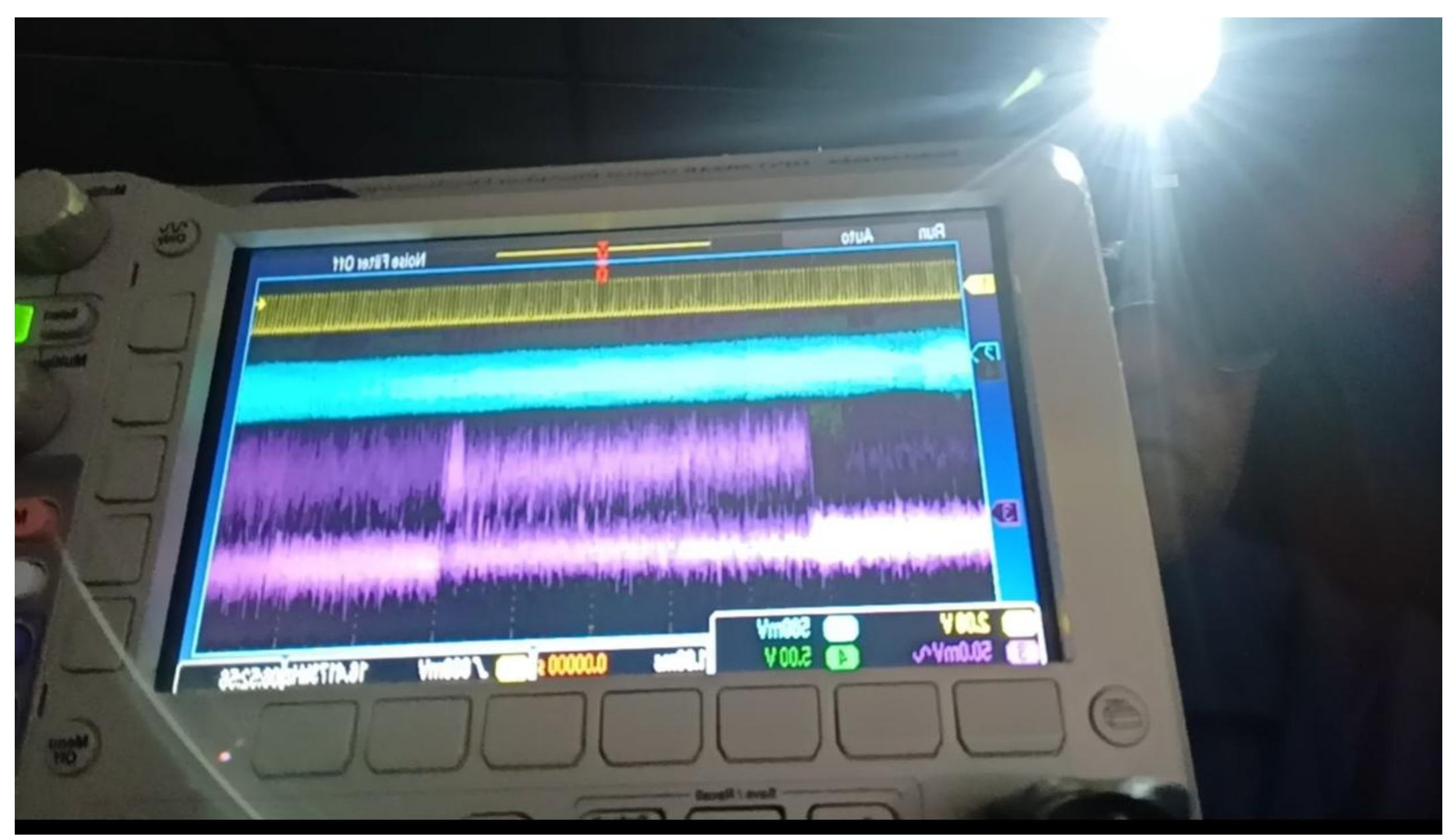

**Figure S2:** Example of quantum information operations performed with our QN-SQ system, where the experimental measurement configuration has an increased duration of laser excitation for the optical spin qubits. The neuron spiking sequence is shown in yellow, and the green/turquoise are the neuron-modulated RF signals in this case. The PL experimental data in magenta now has clear quantum oscillations during the operation of the quantum neurons.